# Crab effect, advantage law, silver medal and other applications of the minority game

J.Menche[1,2] and J.R.L. de Almeida[1]

Departamento de Física, Universidade Federal de Pernambuco [1]
50670-001, Recife, PE, Brazil

Fakultät fur Physik und Geowissenschaften, Universitat Leipzig [2]
LinnestraBe 5, 04103 Leipzig, Germany

PACS numbers: 02.50.Le, 05.40.-a, 64.60.Ak, 89.90.+n

## Abstract

In this work we present a pedagogical introduction to the minority game and various new versions of it with interesting properties, focusing in its applications in socialphysics. For instance, some systems display a kind of social behavior that seems to play an important role in the advancement and survival of an organized society [see, for instance, J. B. Silk et al., Science 302, 1231 (2003)]. On the other hand, devious behavior may degrade a organized society specially when anti-social individual patterns becomes common to many members of a collectivity. In a, perhaps, somewhat far fetched application of a model for interacting agents, the well known minority game, applicable in many contexts, we have studied by computer simulation the effect of having a fraction of the members of a collectivity endowed with spurious strategies. In particular the so called advantage law, where there are some agents that always win, no matter if they play good or not, and another one is a realization of a popularly known "crab effect", where better performing agents may be suppressed by the mass of the medial players. As may be expected, this antisocial strategies deteriorate the collective organization of the system, but now studied within a measurable framework. In another application, positively minded, of the multi choice minority game we introduce different ways to reward also second place winners and compare the results with the one of the standard MG.

Mail addresses: joergmenche@gmx.de ; almeida@df.ufpe.br

# I. INTRODUCTION TO THE MINORITY GAME

The minority game is a very simple model of interacting agents, nevertheless it can be applied to various real situations and shows fascinating properties.

## A. The Standard Minority Game (SMG)

The main idea of the SMG with two choices is the following an odd number of selfish players repeatedly has to choose one out of two possible actions. For example:
- buying or selling at the stock market
- taking the bus or the metro to the university
- 0 or 1 on a computer

All the players take their choice and then we count the players that decided to play '0' and the ones who played '1'. The players that are in the minority side will win the game.

Let's go into the details now. We have $N$(odd!) players in the game, let's say $N = 101$. The only global information about the state of the game lies in a "history" or "memory" of length m of the last $m$ games' winners. There are two possible values (0 or 1) for each of the m entries of the history, so there are $2^m$ possible different histories. Every player is equipped with a set of $S$ strategies to determine their next choice. As the decision depends on the current history of the last games, these strategies contain $2m$ values, each of them representing the choice for one of the possible histories. All of the $2^m$ entries can be 0 or 1, so there are $2^{2^m}$ different strategies. Following we show an example for a history of length $m = 5$.

| | |
|---|---|
| Winner of the last round $r = i$-1 | 0 |
| Winner of round $r = i$-2 | 1 |
| | 1 |
| ….. | 0 |
| Winner of round $r = i$-m | 1 |

Here is an example to illustrate the construction of a strategy for a history of length m = 3.

2 $^m$ possible histories and based on them a decision, what to play if that history occurs

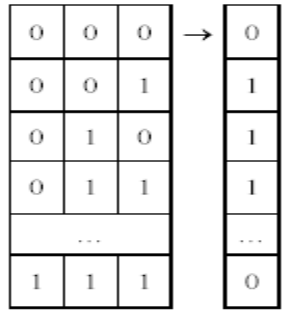

Now we leave our players alone and see what happens. We let them play $R=10000$ rounds and after each round we'll keep the number of players that played 1 $N_i^{(1)}$. As it comes out the standard deviation

$$\sigma\left(N_i^{(1)}\right)= \sqrt{\frac{1}{R}\sum_{i=0}^{R}\left(N_i^{(1)}-\frac{N}{2}\right)^2}$$

is an interesting quantity. It's a measure for the efficiency of the system at distributing the limited resources. If we repeat this game let's say 100 times, we see that depending on the history m the system shows very different behavior. Figures 1 show that the system undergoes a phase transition. In fig.1(b) we scaled various system sizes into a single graph using a $\alpha^2/N$ vs. $2^m/N$ plot, we call $2^m/N = \alpha$. For $\alpha > \alpha_c \approx 0.34$ the system displays a cooperative phase, for small values ($\alpha < \alpha_c$) the coordination of the agents is worse than the one of a "random system". The "random system" consists of agents that make their choice by coin toss, without using any strategy (horizontal line in fig.1b). This behavior can be explained using the formalism of spin glasses. In the cooperative phase the introduced spin variable is frozen, the system becomes predictable.

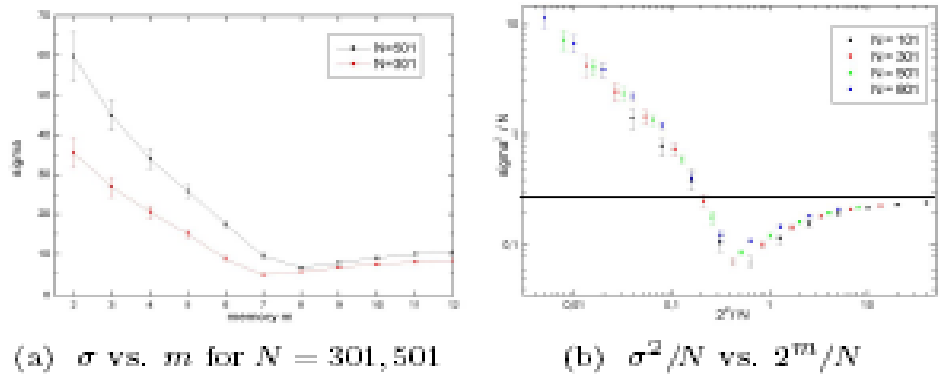

(a) $\sigma$ vs. $m$ for $N = 301, 501$     (b) $\sigma^2/N$ vs. $2^m/N$

FIG. 1: SMG: $S = 2$; $R = 10000$; 100 independent runs; (a) normal plot; (b) scaled plot

**B. The Multi Choice Minority Game (MCMG)**

The introduced Minority Game can be easy generalized to more than two options. The qualitative behavior stays the same (see figure 2), *if we don't change the rules…*

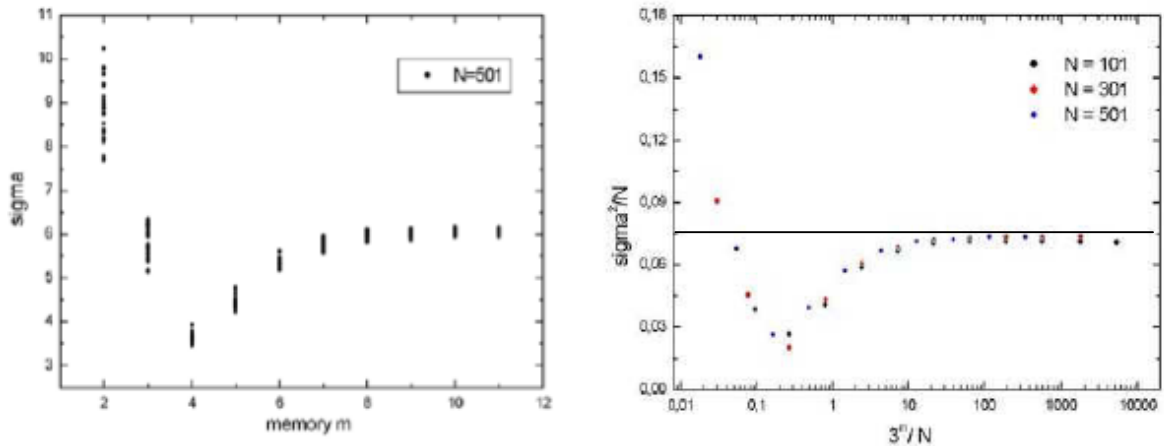

FIG. 2: MCMG: $S = 2$; $R = 10000$; 100 independent runs; (a) normal plot; (b) scaled plot

# II. THE SILVER MEDAL

So let's modify the rules a little and see what happens! Our idea was to make the Minority Game a little bit fairer. As in real life it's sometimes also O.K. to get the second place. W4e found three different ways to implement this in the Minority Game.

**Model 1**

Every agent is equipped with $S$ normal strategies like in the standard Minority Game. We the round is over we will compare our strategies with the result of the game. If a strategy predicted the right result it will be rewarded with $f$ (for first place) points. If the strategy didn't predict a first place, but at least a second place, it will be rewarded with $s$ (for second place). Therefore the maximum amount of points a strategy can win per round is $f$.

**Model 2**

Every agent is equipped with *S* strategies that make *two* predictions at one time, one for the first place and one for the second place. The agents will always use the prediction for the first place to make their choice. If a strategy made a right prediction for the first place, it will be rewarded with *f* points. If it made a right prediction for the second place, it will be rewarded with *s* points. Because the strategies make two predictions they can get *f* + *s* points here. How did we implement this? As in the standard Minority Game the predictions of the strategies are based on the history of the game. So we introduce a separate history $m_2$ for the second places, of the *m* last rounds. The strategies make their prediction for the second place based on this history $m_2$.

**Model 3**

Model 3 is almost the same as Model 2, however, there's a slight difference in the implementation that leads to notable differences in the results:
Instead of introducing an own history $m_2$ for the second places the strategies will make their prediction for the second place based on the history $m_1$. So in this case the plus on information that is given to the players is less.

**A. Results**

Figure 3 shows the results of our simulations. In the uncoordinated phase (m ≥ 4, in fig. 3(a)) there's no difference, all the curves fall on one. For smaller values of *m*, however, there's a big difference.

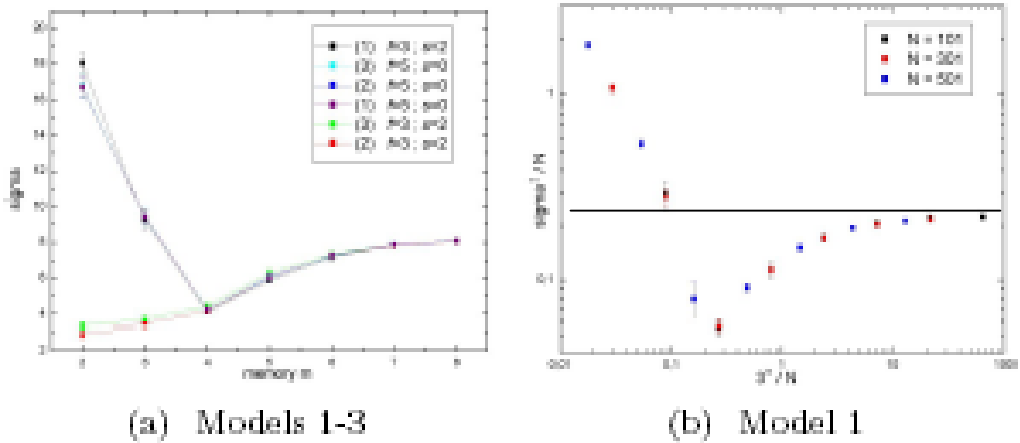

(a) Models 1-3    (b) Model 1

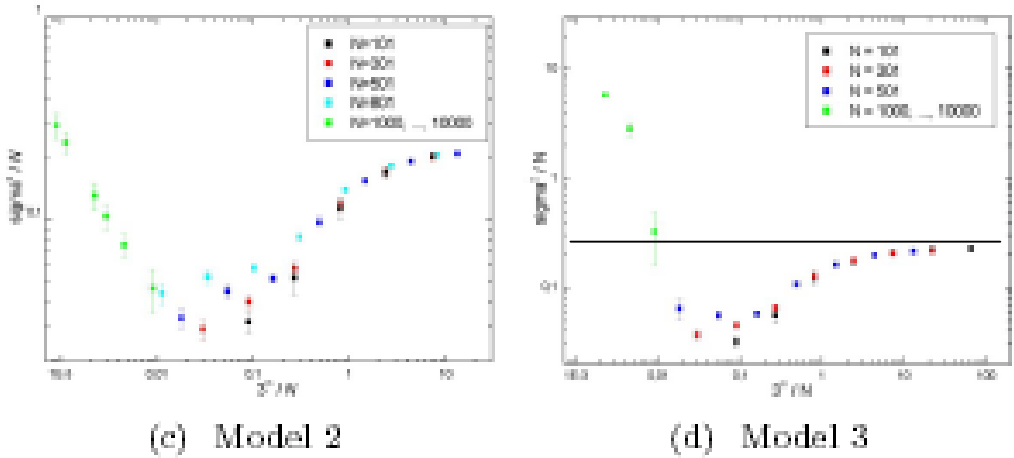

FIG. 3: MCMG: 3(a) shows all the models together in a σ vs.
*m* plot (N = 501)
3(b)-3(d) show the models in a σ²/N vs. $3^m/N$ plot;
parameters: S = 2; R = 10000; 100 independent runs

The results of model 1 are invariant under introducing a reward *s* for the second place. The models 2 and 3 on the contrary show a significantly improved performance for *s* ≠ 0. In these cases the phase transition of the MG is suppressed and the system remains in the correlated phase. We find this very interesting, because it means that the overall performance of these systems is much better, and one may tentatively say "more information available better performance". They show only small fluctuations around the optimal value of $N_{max}$. In the sense of the global win of all agents they are therefore more efficient, because more agents win at the same time. In figures 3(b)-(d) we show the scaled curves of our models. For model 2 we had to go to very high values of *N* to make the phase transition appear (≈10000).

## III. THE ADVANTAGE LAW

The advantage law is a popular brazilian saying (in portuguese lei da vantagem or lei de Gerson, see appendix A), applied whenever someone tries to overcome others at the expenses of their rights. A simple example: overtaking others in a queue. It happens to us every day that somebody tries to take an advantage. There are numerous occasions where some especially "clever" people look only for their own good and therefore break the rules. We introduced this idea into the Minority Game and implemented it in two different ways.

N players are playing the Minority Game as usual. The maximum of points to be distributed is $N_{max} = \frac{N-1}{2}$, they are the resources of the system. Now we introduce an elite that "eats" a part *k* of the available resources. We implemented two realizations of the idea.

**Model 1**

If the total number of winners $N_{win}$ plus the $k$ "clever" players exceeds the for everyone available resources ($N_{win} + k > N_{max}$) the members of the group $N_{win}$ become looser.

**Model 2**

If the total number of winners $N_{win} + k$ exceeds the available resources ($N_{win} + k > N_{max}$) the agent's strategies will only win $\frac{N_{max}}{N^{win} + k} < 1$ points, instead of 1.

### A. Results

Figure 4(a) shows the standard deviation vs. memory for model 1 for $k = 0, 4, 6,$ and 10 which for the number of agents considered is approximately the percentage of "clever" agents in the system. Clearly the overall performance of the adaptive system becomes worse as soon as $k$ is different from zero. By depriving the normal players of their reward the extra agents increase the fluctuations in the number of agents in the minority side thus favoring higher fluctuations and this happens for any concentration of the extra players for model 1. At about 10% of their presence the averaged standard fluctuation is always above the random coin tossing case (horizontal line in the graph) and there is no longer a optimum distribution (minima for σ).

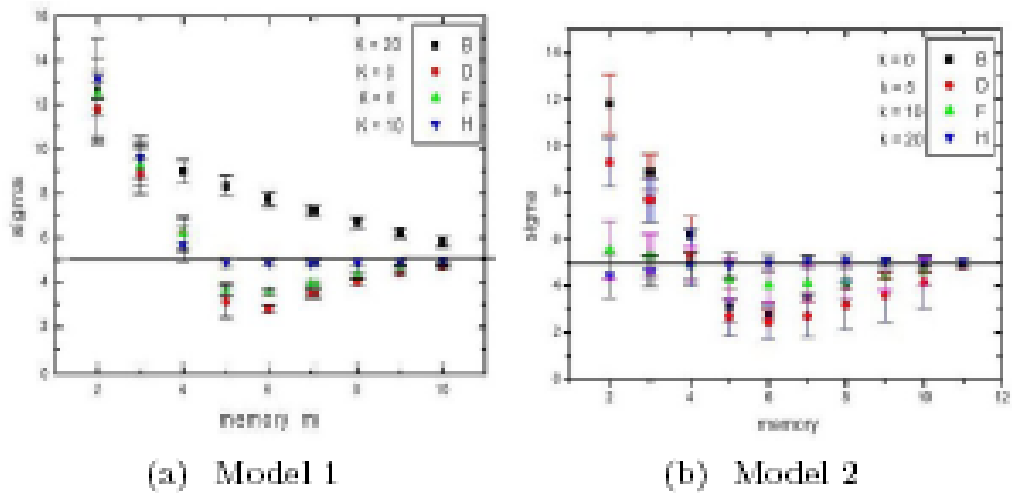

(a) Model 1  (b) Model 2

FIG. 4: Advantage Law: parameters $S = 2$; $R = 10000$; 100 independent runs;

Figure 4(b) shows the standard deviation vs. memory for model 2 for $k = 0, 5, 10$ and 20. Clearly the overall performance of the adaptive system is again worse as soon as $k$ is different from zero and again depriving the normal players of their full reward, the extra agents increase the fluctuations in the number of agents in the minority side thus favoring higher fluctuations, ie., a large spread in $\sigma$, and this happens for any concentration of the extra players for model 2. In this case for low memories their presence may improves somewhat the performance (which always a bad one) but at high memory values for all $k$ considered the system performs worse than the random coin case.

## IV. THE CRAB EFFECT

The crab effect is also a popular brazilian saying (see appendix B to see how Chico Science describes the crab effect in two of his songs). A society that pulls the successful down, just as the crabs in a cooking pot. Some crabs manage to reach the top and almost succeed to get out of the boiling water but the other crabs pull them back... Some individuals disorganize the whole system in order to achieve personal advantages.

We implemented the crab effect as follows: during the game the strategies of the players $i$ gain points $P^s_i$. Better players have strategies with more points. So we simply cut the points of the strategies, when they exceed a certain limit. This limit is set by the medium $P_{medium}$ of all the player's strategies. If a strategy has more points than the average of all strategies multiplied by a parameter c, the points of this strategy will be reset to $P_{medium}$.

$$P^s_i > c \cdot P_{medium} \Rightarrow P^s_i = P_{medium}$$

### A. Results

Figure 5(a) shows our results in an overview, whereas figure 5(b) gives more detailed informations in a 3D plot. Clearly the presence of "crab-like" behavior leads to disorganization in the system. For small values of c, which corresponds to a strong suppression of successful players the system becomes a system worse than that of pure random choice.

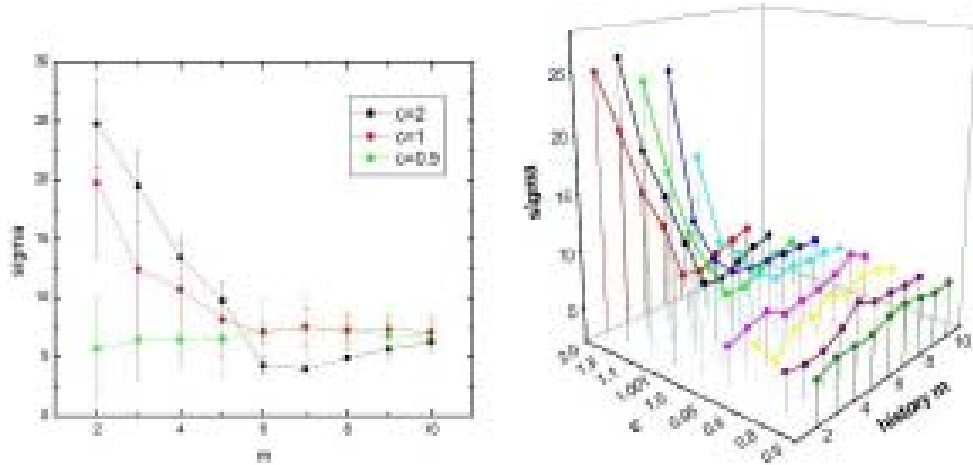

FIG. 5: Crab effect: parameters $S = 2$; $R = 10000$; 100 independent runs;

## V. CLOSING REMARKS

In this paper we gave a very short introduction to the Minority Game and presented some new versions of it. However the Minority Game has a lot more to offer. It asks challenging questions to statistical mechanics and its methods and ideas. Some say that it may come to play the same role in science as the Ising model, is applicable in many interdisciplinary subjects from physics to economy and biology to name a few.

The aim of this paper was basically to make you curious and to introduce you into the fascinating properties of the Minority Game. There are many things to be done in the area of "pure" physics as well as at the frontier to other disciplines. The best source for comprehensive and actual information is the website http://www.unifr.ch/econophysics, where the ever growing Minority Game community meats. For further information take a look at the references, where we listed some relevant papers and do not hesitate to write us an email.

[1] J. B. Silk et al, Science 302, 1231 (2003).
[2] http://www.aglioeolio.hpg.ig.com.br/cr006.htm
[3] http://www.unifr.ch/econophysics.
[4] W. B. Arthur, Amer. Econ. Review 84, 406 (1994).
[5] R. Metzler, Ph.D. thesis, Universität Würzburg (2002).
[6] Y.-C. Zhang, Europhys. News 29, 51 (1998).
[7] D. Chalet and Y.-C. Zhang, Physica A 246, 407 (1997).
[8] M. Sysi-Aho, A. Chakraborti, and K. Kaski (2003), condmat/0305283v1.
[9] D. Challet and M. Marsili, Phys. Rev. E 60, R6271 (1999).
[10] L. Ein-Dor, R. Metzler, I. Kanter, and W. Kinzel, Phys. Rev. E 63, 066103 (2001).
[11] R. Savit, R. Manuca, and R. Riolo, Phys. Rev. Let. 82, 2203 (1998).
[12] D. Challet, M. Marsili, and R. Zecchina, Phys. Rev. Let. 84, 1824 (2000).

# APPENDIX A: A LEI DA VANTAGEM [2]

Gerson, um dos grandes craques daquela espetacular seleção de 1970, a do Tri, teve a infelicidade de gravar um comercial para uma marca de cigarro que prometia aquelas maravilhas de sempre. O nosso "canhotinha de ouro", como bom fumante, encerrava o comercial pronunciando, com aquele sotaque bem carioca, a frase: "gosto de levar vantagem em tudo, certo?"

A maldição do comercial estigmatizou para sempre o nosso craque. "Gosto de levar vantagem em tudo" ficou célebre e virou slogan. Estava criada a Lei de Gerson, que serve at´e hoje para justificar os pequenos e grandes delitos que são cometidos diariamente em nosso país.

<div style="text-align:center">Gosto de levar vantagem em tudo, certo? ERRADO !!!</div>

Quem nunca viu ou ficou fulo com essas situações?
- O malandro que corta a fila no banco quando vê um amigo lá na frente
- O "mano" passando por baixo da roleta do ônibus (nunca vi integrante de torcida organizada pagar passagem)
- Centrais de Telemarketing que coletam doações para casas de crianças com câncer, velhinhos abandonados, etc; e que na verdade são estelionatários
- E, do que temos mais medo, malandros que roubam e seqüestram pessoas, confiando na incapacidade da polícia e dos tribunais de investigar e puni-los, talvez com garantias (corrupção e troca de favores) ou mesmo sem elas.

Acho que vale refletir um pouco sobre o assunto. Nossos políticos são denunciados diariamente pelo comportamento antiético no exercício da função pública. São dezenas e dezenas de casos escabrosos que chegam ao conhecimento das pessoas. Na maioria das vezes, nada acontece. Fica por isso mesmo e tudo acaba em pizza. Algumas CPIs são criadas, para a aparição de "Xerifes" e bodes-expiatórios têm prisão provisória, mas a imagem dos políticos, sindicalistas e policiais corruptos não se abala. Alguns poucos resultam em cassações de mandatos e retirada de candidaturas. E só. As fortunas construídas ilegalmente continuam nas mãos (ou em contas no exterior) dos donos. Ou seja, levaram vantagem a vida toda e agora, com dinheiro e poder (e parte da mídia) podem gozar a vida graças a impunidade nossa de cada dia.

Policiais mal pagos que aceitam propina de traficantes poderosos, Fiscais que têm sérios problemas de vista e ficam ricos de uma hora pra outra, a Corrupção talvez seja o pior "produto de Exportação" de nosso país, e ela se deve (e muito) a essa condição de impunidade ao "Malandro", onde os trouxas somos nós, mas talvez por corporativismo da "malandragem" ninguém se sinta ofendido. Afinal, O Coronel do Massacre de Carandiru, Euvírus Miranda, ACM, José Sarney, Garotinho e outros Coronéis são ídolos da "Massa" e campeões de votos.

Essa cultura da malandragem e da esperteza está enraizada no nosso dia a dia. E é até incentivada na TV. O apresentador Pedro Bial, comentando o lance do pênalti no atacante Luizão na Copa 2002, em rede nacional, afirmou que "roubado é mais gostoso". Uma frase infeliz, um mau exemplo para a formação da garotada. [. . .]

**taken from: http://www.aglioeolio.hpg.ig.com.br/cr006.htm**

# APPENDIX B: CHICO SCIENCE AND THE CRAB EFFECT

Chico Science

| Da lama ao caos | A Cidade |
|---|---|
| Posso sair daqui para me organizar | O sol nasce e ilumina as pedras evoluídas |
| Posso sair daqui para desorganizar | Que cresceram com a força de pedreiros suicidas |
| Da lama ao caos, do caos a lama | Cavaleiros circulam vigiando as pessoas |
| Um homem roubado nunca se engana | Não importa se são ruins, nem importa se são boas |
| O sol queimou, queimou a lama do rio | E a cidade se apresenta centro das ambições |
| Eu ví um Chié andando devagar | Para mendigos ou ricos e outras armações |
| Ví um aratu pra lá e pra cá | Coletivos, automóveis, motos e metrôs |
| Ví um carangueijo andando pro sul | Trabalhadores, patrões, policiais, camelôs |
| Saiu do mangue, virou gabiru | A cidade não para, a cidade só cresce |
| Oh Josué eu nunca ví tamanha desgraça | O de cima sobe e o de baixo desce |
| Quanto mais miséria tem, mais urubu ameaça | A cidade não para, a cidade só cresce |
| Peguei o taláio, fui na feira roubar tomate e cebola | O de cima sobe e o de baixo desce |
| Ia passando uma véia, pegou a minha cenoura | A cidade se encontra prostituída |
| Aí minha véia, deixa a cenoura aquí | por aqueles que a usaram em busca de saída |
| Com a barriga vazia não consigo dormir | Ilusora de pessoas de outros lugares |
| E com o bucho mais cheio começei a pensar | a cidade e sua fama vai além dos mares |
| Que eu me organizando posso desorganizar | No meio da esperteza internacional |
| Que eu desorganizando posso me organizar | A cidade até que não está tão mal |
| Da lama ao caos | E a situação sempre mais ou menos |
| Do caos à lama | Sempre uns com mais e outros com menos |
| Um homem roubado, nunca se engana | A cidade não para, a cidade só cresce |
| | O de cima sobe e o de baixo desce |
| | A cidade não para, a cidade só cresce |
| | O de cima sobe e o de baixo desce |
| | Eu vou fazer uma embolada, um samba, um maracatu |
| | Tudo bem envenenado, bom pra mim e bom pra tu |
| | Pra a gente sair da lama e enfrentar os urubu |
| | Num dia de sol Recife acordou |
| | Com a mesma fedentina do dia anterior. |